\documentclass[prd,twocolumn,superscriptaddress,preprintnumbers,showpacs,
nofootinbib]{revtex4}
\usepackage{graphicx}
\usepackage{epsfig}
\usepackage{bm}
\usepackage{latexsym,amssymb,amsmath,amsfonts,amssymb,wasysym,float}
\usepackage{scrextend}
\usepackage{mathrsfs}
\usepackage{color}

\usepackage{enumitem}

\newcommand{\postscript}[2]{\setlength{\epsfxsize}{#2\hsize}
   \centerline{\epsfbox{#1}}}

\usepackage[usenames,dvipsnames]{xcolor}
\definecolor{orange}{cmyk}{0,0.5,1,0}
\definecolor{rossoCP3}{cmyk}{0,.88,.77,.40}
\definecolor{graa}{rgb}{0.8,0.8,0.8}
\definecolor{blaa}{rgb}{0.2,0.2,0.6}

\begin{document}
\preprint{MPP-2023-139}
\preprint{LMU-ASC 24/23}

\title{\color{rossoCP3} Fuzzy Dark Matter and the Dark Dimension}

\author{\bf Luis A. Anchordoqui}

\affiliation{Department of Physics and Astronomy,  Lehman College, City University of
  New York, NY 10468, USA
}

\affiliation{Department of Physics,
 Graduate Center, City University
  of New York,  NY 10016, USA
}

\affiliation{Department of Astrophysics,
 American Museum of Natural History, NY
 10024, USA
}

\author{\bf Ignatios Antoniadis}
\affiliation{Laboratoire de Physique Th\'eorique et Hautes \'Energies - LPTHE
Sorbonne Universit\'e, CNRS, 4 Place Jussieu, 75005 Paris, France
}

\author{\bf Dieter\nolinebreak~L\"ust}

\affiliation{Max--Planck--Institut f\"ur Physik,  
 Werner--Heisenberg--Institut,
80805 M\"unchen, Germany
}

\affiliation{Arnold Sommerfeld Center for Theoretical Physics, 
Ludwig-Maximilians-Universit\"at M\"unchen,
80333 M\"unchen, Germany
}

\begin{abstract}
  \vskip 2mm \noindent We propose a new dark matter contender within
  the context of the so-called ``dark dimension'', an innovative
  5-dimensional construct that has a compact space with characteristic
  length-scale in the micron range. The new dark matter candidate is
  the radion, a bulk scalar field whose quintessence-like potential
  drives an inflationary phase described by a 5-dimensional de Sitter
  (or approximate) solution of Einstein equations. We show that the
  radion could be ultralight and thereby serve as a fuzzy dark matter
  candidate. We advocate a simple cosmological production mechanism
  bringing into play unstable Kaluza-Klein graviton towers which are
  fueled by the decay of the inflaton. 
\end{abstract}
\date{July 2023}
\maketitle

\section{Introduction}

While there are various lines of evidence for the existence of dark
matter in the universe, the nature of the dark matter particle
remains a challenging dilemma at the interface of astrophysics,
cosmology, and particle physics~\cite{Feng:2010gw}. There is a large variety of dark matter candidates with masses
spanning many orders of magnitude. Of particular interest here, fuzzy
dark matter (FDM) is made up of non-interacting ultralight bosonic
particles that exhibit coherent dynamics and a wave-like behaviour on
galactic scales~\cite{Hu:2000ke}. On sub-galactic length scales, FDM brings to light a
distinctive phenomenology alternative to that of cold dark matter
(CDM). However, FDM predictions are indistinguishable from those of
CDM on large scales, and so benefits from the remarkable success of
$\Lambda$CDM cosmology.

The main parameter regulating the two FDM regimes is the particle's
mass, with a range that spans three decades of energy, $10^{-24} \alt m/{\rm eV} \alt
10^{-22}~{\rm eV}$. Particles of such tiny
mass and with typical velocities $v$ found in haloes hosting Milky
Way-sized galaxies, acquire a very long
de Broglie wavelength,
\begin{equation}
\lambda_{\rm dB} \equiv \frac{2 \pi}{mv} = 4.8~{\rm kpc}
\left(\frac{10^{-23}~{\rm eV} }{m}\right) \left(\frac{250~{\rm
      km/s}}{v}\right) \,, 
\end{equation}
to deliver the wave-like behavior at galactic scales.

FDM can populate the galactic haloes with large occupation numbers and
behave as self-gravitating dark matter waves. This engenders a pressure-like effect on macroscopic scales which catalyzes a flat core at the center of galaxies, with a
relatively marked transition to a less dense outer region that follows
the typical CDM-like distribution. 

Before proceeding, we take note of a serious challenge for FDM models. The
criticism centers on numerical simulations to accommodate
Lyman-$\alpha$ forest data which provide bounds on the fraction of 
FDM~\cite{Irsic:2017yje,Armengaud:2017nkf,Rogers:2020ltq}. In
response, it was noted that these bounds strongly depend on the modeling of
the intergalactic medium~\cite{Hui:2016ltb}. More recently, new
constraints for FDM models have emerged, e.g. {\it (i)}~from inferences of the
low-mass end of the subhalo mass function~\cite{Schutz:2020jox}, {\it (ii)}~from
observations of ultra-faint dwarf galaxies~\cite{Dalal:2022rmp}, and
{\it (iii)}~from 
superradiance of FDM which would cause the supermassive black hole at
the center of M87 to spin down excessively~\cite{Davoudiasl:2019nlo}. Constraints {\it (i)} and
{\it (ii)} also depend on simulations and are subject to a completely
different set of assumptions and systematic
uncertainties. The Event Horizon Telescope measurement of the spin of
M87$^*$ excludes FDM masses in the range $10^{-21} \alt m/{\rm eV} \alt 10^{-20}$. Whichever point of view
one may find more convincing, it seems most conservative at this point
to depend on experiment (if possible) rather than numerical simulations to resolve the issue.

In this paper we show that the dark dimension
scenario~\cite{Montero:2022prj} embraces a well motivated FDM
candidate. The
layout is as follows: in Sec.~\ref{sec:2} we outline the basic setting of the dark
dimension scenario and we identify the radion as a FDM candidate, in Sec.~\ref{sec:3} we discuss the
process of radion production to estimate the corresponding relic
abundance, and conclusions are given in Sec.~\ref{sec:4}.

\section{The Good, the Bad, and the Fuzzy} 
\label{sec:2}

The Swampland  program seeks to understand
which are the ``good'' low-energy effective field theories (EFTs) that
can couple to gravity consistently (e.g. the landscape
of superstring theory vacua) and distinguish them from the
``bad'' ones that cannot~\cite{Vafa:2005ui}. In theory space, the
frontier discerning the
good theories from those downgraded to the swampland is drawn
by a family of conjectures classifying the properties that an EFT should
call for/avoid to enable a consistent completion into quantum gravity. These
conjectures provide a bridge from quantum gravity to astrophysics,
cosmology, and particle physics~\cite{Palti:2019pca,vanBeest:2021lhn,Agmon:2022thq}.

For example, the distance conjecture (DC) forecasts the appearance of infinite towers of states that become
exponentially light and trigger the collapse of the EFT at infinite distance limits
in moduli space~\cite{Ooguri:2006in}. Connected to the DC is the anti-de
Sitter (AdS) distance conjecture, which
correlates the dark energy density to the mass scale $m$ characterizing the infinite tower of states,
$m \sim |\Lambda|^\alpha$, as the negative AdS vacuum energy
 $\Lambda \to 0$, with $\alpha$ a positive constant of ${\cal O}
 (1)$~\cite{Lust:2019zwm}. Besides, under the hypothesis that
 this scaling behavior holds in dS (or
 quasi dS) space, an unbounded number of massless modes also pop up in the limit $\Lambda \to 0$.

As demonstrated in~\cite{Montero:2022prj}, the generalization of the AdS-DC to dS space could help
elucidate the radiative stability of the cosmological hierarchy $\Lambda_4/M_p^{4} \sim 10^{-120}$, because it connects the
size of the compact space $R_\perp$ to the
dark energy scale $\Lambda_4^{-1/4}$,
\begin{equation}
  R_\perp \sim \lambda \ \Lambda_4^{-1/4} \,,
\label{Rperp}  
\end{equation}
where the proportionality factor is estimated to be within the range
$10^{-1} < \lambda < 10^{-4}$. Actually, (\ref{Rperp}) derives from
  constraints by theory and experiment. On the one hand, since the associated Kaluza-Klein (KK) 
  tower contains massive spin-two bosons, the Higuchi
  bound~\cite{Higuchi:1986py}  provides an absolute upper limit to
  $\alpha$, whereas explicit string calculations
of the vacuum energy~(see
e.g.~\cite{Itoyama:1986ei,Itoyama:1987rc,Antoniadis:1991kh,Bonnefoy:2018tcp})
yield a lower bound on $\alpha$. All in all, the theoretical
constraints lead to $1/4 \leq \alpha \leq 1/2$; see~\cite{Anchordoqui:2023laz} for a
recent discussion. On the other hand, experimental arguments (e.g. constraints on deviations from Newton's gravitational inverse-square law~\cite{Lee:2020zjt} and neutron star heating~\cite{Hannestad:2003yd}) lead to the conclusion encapsulated in (\ref{Rperp}); namely, that there is one extra dimension of
radius $R_\perp$ in the micron range, and that the lower bound for $\alpha =
1/4$ is basically saturated~\cite{Montero:2022prj}. This in turn
implies that the KK tower of the new (dark)
dimension opens up at the mass scale $m_{\rm KK} \sim
1/R_\perp$. Within this set-up, the 5-dimensional Planck scale (or species scale
where gravity becomes strong~\cite{Dvali:2007hz,Dvali:2007wp}) is
given by $M_* \sim m_{\rm KK}^{1/3}
M_p^{2/3}$. Note that for $m_{\rm KK} \sim 1~{\rm eV}$, we have $M_* \sim
  10^{9}~{\rm GeV}$ and therefore the species scale is outside the reach of collider experiments~\cite{ParticleDataGroup:2022pth}.

The dark dimension stores a top-notch   
phenomenology~\cite{Anchordoqui:2022ejw,Anchordoqui:2022txe,Blumenhagen:2022zzw,Anchordoqui:2022tgp,Gonzalo:2022jac,Anchordoqui:2022svl,Anchordoqui:2023oqm,vandeHeisteeg:2023uxj, Noble:2023mfw,Anchordoqui:2023wkm,Anchordoqui:2023qxv}. For
example, it was noted in~\cite{Gonzalo:2022jac} that the universal coupling of the SM
fields to the massive spin-2 KK excitations of the graviton in the
dark dimension provides a dark matter candidate. Complementary to the
dark gravitons, it was observed in~\cite{Anchordoqui:2022txe} that primordial black holes
with Schwarzschild radius smaller than a micron could also be good
dark matter candidates, possibly even with an interesting close
relation to the dark gravitons~\cite{Anchordoqui:2022tgp}. Next, in line with our stated plan,  we propose a new
dark matter candidate within this framework.

It is unnatural to entertain that the size of the dark dimension would
remain fixed during the evolution of the Universe right at the species
scale.  To accommodate this hierarchy we need to inflate the size of
the dark dimension. To see how this works explicitly, we consider that the inflationary phase can
described by a 5-dimensional dS (or approximate) solution of Einstein equations~\cite{Anchordoqui:2022svl}. All dimensions (compact and non-compact)  expand
exponentially in terms of the 5-dimensional proper time. This implies
that when inflation starts the radius $R$ of the compact
space is small and the
4-dimensional Planck mass is of order the 5-dimensional Planck scale $M_*$. However, when inflation ends the radius of the compact
space is on the micron-scale size and the 4-dimensional Planck scale is much
bigger, 
\begin{equation}
  M_p^2 = 2 \pi \ M_*^3 \ R_\perp \, .
\label{mpms}
\end{equation}
A straightforward calculation shows that the compact space requires 42
e-folds to expand from the fundamental length $1/M_*$ to the micron
size. We can interpret the solution in terms of 4-dimensional  
fields using 4-dimensional Planck units from the relation (\ref{mpms}), which
amounts going to the 4-dimensional Einstein frame. 
  Namely, the higher-dimensional metric in $M_*$ units is given by
\begin{equation}
  ds_5^2 =  a_5^2 \ (-d\eta^2 + d\vec x^2 + r^2_0 \ dy^2) \,,
\label{metric}
\end{equation}
where $\eta$ is the conformal time, $a_5 = 1/(H\eta)$, $H$ is the
Hubble parameter, $\vec x$ denotes the 3 uncompactified
dimensions, and $r_0 \sim 1$ is the radius of the
dark dimension $y$ at the beginning of the inflationary phase. The
4-dimensional decomposition in the Einstein frame is given by
\begin{equation}
  ds_5^2 = \frac{1}{R} ds_4^2 + R^2 dy^2 \,,
\label{metric2}  
\end{equation}  
where $ds_4^2 = a_4^2 (-d\eta^2 + d^2\vec x)$. Comparing
(\ref{metric}) and (\ref{metric2}) we arrive at $a_4/\sqrt{R} = R$. After inflation of $N$ $e$-folds,
where the scale factor was expanded by $a_5 = e^N$, the radius becomes
$R=e^N$.  This implies that if $R$ expands
$N$ e-folds, then the 3-dimensional space would expand $3N/2$ e-folds as a result of
a uniform 5-dimensional inflation~\cite{Anchordoqui:2022svl}. We want
$r_0$ to grow fast up to the micron scale. Altogether, the 3-dimensional space has expanded by about
60 e-folds to solve the horizon problem, while connecting this
particular solution to the generation of a mesoscopic size dimension. A consistent model requires the size of the dark dimension to be
stabilized at the end of inflation; an investigation along this line
is already presented in~\cite{Anchordoqui:2023etp}.

The 5-dimensional action of uniform dS (or approximate) inflation 
\begin{equation}
S_5=\int[d^4x][dy] \left({1\over 2} M_*^3 {\cal
    R}_{(5)}-\Lambda_{5}\right) \,,
\end{equation}
leads to a runaway potential for the radion $R$ coming from the
5-dimensional cosmological constant, where ${\cal
  R}_{(5)}$ is the higher
dimensional curvature scalar and $\Lambda_5$ is the 5-dimensional cosmological
constant at the end of inflation. The quintessence-like potential of the
radion is seen explicitly upon dimensional reduction to 4 
dimensions. The resulting 4-dimensional action in the
Einstein frame is found to be,
\begin{eqnarray}
S_{4} &  = &  \int[d^4x] \left\{{1\over 2} \ M_p^{2} \ {\cal R}_{(4)} -
             {3\over
    4} \ M_p^2 \ \left({\partial R\over R}\right)^2 \right. \nonumber \\
  & - & \left. (2\pi \langle R \rangle)^{2}  {\Lambda_{5}\over (2\pi R)}\right\}\,,  
\end{eqnarray}
where ${\cal R}_{(4)}$ is the Ricci scalar and $\langle R \rangle$ is
the vacuum expectation value (vev) of $R$ after the end of inflation. Because the radion field $R$ is not canonically normalized, we define
$\phi = \sqrt{3/2} \ \ln \left(R/\langle R \rangle \right)$. In terms
  of the normalized field $\phi$ the scalar potential takes the
  advertised quintessence-like form 
  \begin{equation}
    V(\phi) = 2 \pi \ \Lambda_5 \ \langle R \rangle \ e^{- \sqrt {2/3}
      \phi} \, .
\label{Vphi}
\end{equation}
Exponential
potentials of the form $e^{-\alpha \phi}$ are constrained by
cosmological and astrophysical 
observations. The existing data lead to an upper bound 
$\alpha \alt
0.8$~\cite{Barreiro:1999zs}. Curiously, the upper limit on the allowed
value of $\alpha$ is the one predicted by (\ref{Vphi}).\footnote{This
 exponent is also compatible with the dS swampland conjecture~\cite{Obied:2018sgi}.} Even though the potential in (\ref{Vphi}) could be used to explain the current acceleration of the universe, herein we consider the possibility that the
radion is stabilized by additional terms in the potential. The mass of
the radion,
\begin{equation}
  m \sim \sqrt{V^{\rm tot}_{\phi\phi} (0)}/M_p \, ,
\label{meq}
\end{equation}
depends on the functional form of the various additional 
terms $V_i(\phi)$ that allow minimization of the potential, i.e.
\begin{equation}
  V^{\rm tot} (\phi) = V(\phi) + \sum_i V_i(\phi) + \Lambda_4
\end{equation}  
with $V_\phi^{\rm tot} (0) = 0$ and where $V_\phi \equiv dV/d\phi$. Adding
only the term originating from the Casimir energy~\cite{Arkani-Hamed:2007ryu} leads to a lower bound
on $m \sim \sqrt{\Lambda_4}/M_p \sim 10^{-30}~{\rm
  eV}$~\cite{Anchordoqui:2023etp}. 

An important aspect of this model is that the coupling of the radion
to SM fields must be suppressed to avoid conflicts with limits on long
range forces. Herein, we assume that the radion has a localized
kinetic term through (e.g. an expectation value of a brane field) that
suppresses the coupling to matter. Alternatively, in the absence of a scalar potential, the 5-dimensional radion is equivalent to a Brans-Dicke scalar with a parameter $\omega = -4/3$. It has been argued that an appropriate modification of such theories due to bulk quantum corrections can lead to a logarithmic scale (time) dependence of $\omega$ that suppresses the radion coupling to matter, consistently with the experimental limits~\cite{Albrecht:2001cp}. An investigation along these lines is obviously important to be done.

\section{Production of the fuzzy radion and its relic abundance}
\label{sec:3}

The issue that remains to be assessed is whether there is a mechanism
which allows enough radion production to  accommodate the relic
dark matter density. An interesting possibility emerges if the inflaton has roughly equal couplings to
brane and bulk fields, such that on decay will produce the SM fields
while also populating the KK towers.

We begin by considering a tower of equally spaced dark gravitons,
indexed by an integer $l$, and mass  $m_l = l  m_{\rm KK}$. We assume that the cosmic evolution of
the dark sector is mostly driven by ``dark-to-dark'' decay processes
that regulate the decay of KK gravitons within the dark tower. The
proposed decay model then provides a particular realization of the dynamical dark
matter model~\cite{Dienes:2011ja}. The intra-KK decays in the bulk require a
spontaneous breakdown of the translational invariance in the compact
space such that the 5-dimensional momenta are not conserved. An
explicit realization of this idea, in which the KK modes acquire
a nonzero vev $\langle \varphi_l \rangle$ has been
given in~\cite{Mohapatra:2003ah}. Following~\cite{Dienes:2008qi}, we further assume transitions by instanton-induced tunneling dynamics associated
with such vacuum towers. The effect of the instanton processes is to
accelerate the cascade dynamics to collapse into the
radion.

Bearing this in mind, we calculate the decay of a given massive KK graviton
into the radion $\phi$ and a lighter KK graviton in the presence of an
expectation value for the
bulk scalar $\varphi$ that breaks momentum conservation.
Following~\cite{Mohapatra:2003ah} we postulate the existence of a coupling of the form
\begin{equation}
\mathscr{L}_I \supset \ \lambda \ \frac{1}{M_*} \ \Phi \
  h_{AB} \ h_{CD} \ h_{EF} \ {\cal C}^{ABCDEF} \,,
\label{calL}
\end{equation}
coming from 
\begin{equation}
\int d^4x \ dy \ \sqrt{g} \ T (x,y) \,, 
\end{equation}
where $T (x,y)$ is the trace of the energy
momentum tensor of the bulk theory, $h_{AB}$ is the
5-dimensional graviton which comes from expansion of the 5-dimensional
metric around flat space, $g_{AB} = \eta_{AB} +M_*^{-3/2} h_{AB}$, and
where in (\ref{calL}) $\lambda$ is a dimensionless coupling and ${\cal C}^{ABCDEF}$ is a
constant tensor. Next, we expand all 5-dimensional fields in terms of
their 4-dimensional KK modes, for instance
\begin{equation}
\Phi (x,y) = \sum _{l} \frac{1}{\sqrt{2 \pi R_\perp}} \varphi_{l} (x) \
e^{i l y/R_\perp} \, .
\end{equation}  
After integration over $dy$,  (\ref{calL}) can be recast as
\begin{equation}
\mathscr{L}_I \supset \sum_{l,l'} \lambda \ \left(\frac{M_*}{M_p} \right)^2  \
\varphi_{(l-l')} \ \phi \
  h_{l} \ h_{l'} \ {\cal C} \, ,
\end{equation}
where we have made use of
(\ref{mpms}) and considered a factor of $1/\sqrt{2\pi R_\perp}$ for
each KK decomposition of the four fields
 and a factor of $2 \pi R_\perp$ from the integration over
$y$. Now, following~\cite{Mohapatra:2003ah} we assume that $\varphi_l$  takes a vev
which is independent of $l$. The total decay width of a KK
graviton of mass $m_l = l m_{\rm KK}$ is found to be~\cite{ParticleDataGroup:2022pth}
\begin{eqnarray}
\Gamma^l_{\rm tot} & = &  \frac{\lambda^2}{8\pi} \ \frac{1}{m_l^2} \sum_{l' < l} 
\left(\frac{M_*}{M_p}\right)^4 \frac{m_l^2-m_{l'}^2}{2 m_l} \langle \varphi_{l-l'}
                         \rangle^2 \nonumber \\
& = & \frac{\lambda^2}{8 \pi} \  \frac{1}{m_l^2} \
      \left(\frac{M_*}{M_p}\right)^4 \ \frac{(l-1) (4l+1)}{12} m_{\rm KK} \ \langle
      \varphi \rangle^2 \nonumber \\
  & = & \frac{\lambda^2}{96 \pi^3} \left(m_{\rm KK} - \frac{3 m_{\rm
        KK}^2}{4m_l} - \frac{m_{\rm KK}^3}{4 m_l^2} \right)  \left(\frac{\langle \varphi
        \rangle}{M_*} \right)^2 \!\!\!\!,
 \label{KKRfinal}       
\end{eqnarray}
where $\langle \varphi_{(l-l')} \rangle = \langle
\varphi \rangle$. Substituting in (\ref{KKRfinal}) our fiducial value
for $m_{\rm KK} \sim
10~{\rm eV}$ while considering as entertained
in~\cite{Mohapatra:2003ah}  $\langle \varphi \rangle \sim M_*$ we obtain a total decay
width of $\Gamma^l_{\rm tot} \sim 5 \times 10^{12}~{\rm s}^{-1}$,
where we have taken $\lambda \sim 1$ and $m_l\gg m_{\rm KK}$. If we instead
adopt $\langle \varphi \rangle \sim 5 \times 10^{-4} M_*$ we obtain
$\Gamma^l_{\rm tot}  \sim  10^{6}~{\rm s}^{-1}$, which implies
that the energy the inflaton deposited in the KK tower ends
up into the radion well before the QCD phase transition (with
characteristic temperature $\sim 150~{\rm MeV}$ and age $\sim
20~\mu{\rm s}$). Altogether, we conclude that even for $\langle
\varphi \rangle \ll
M_*$ the energy the inflaton
deposited in the KK tower collapses all into the radion well before
the earliest observational verified landmark ({\it viz.}, big bang
nucleosynthesis 
with starting age of roughly 180~s).

\begin{figure*}[htpb!]
  \begin{minipage}[t]{0.49\textwidth}
  \postscript{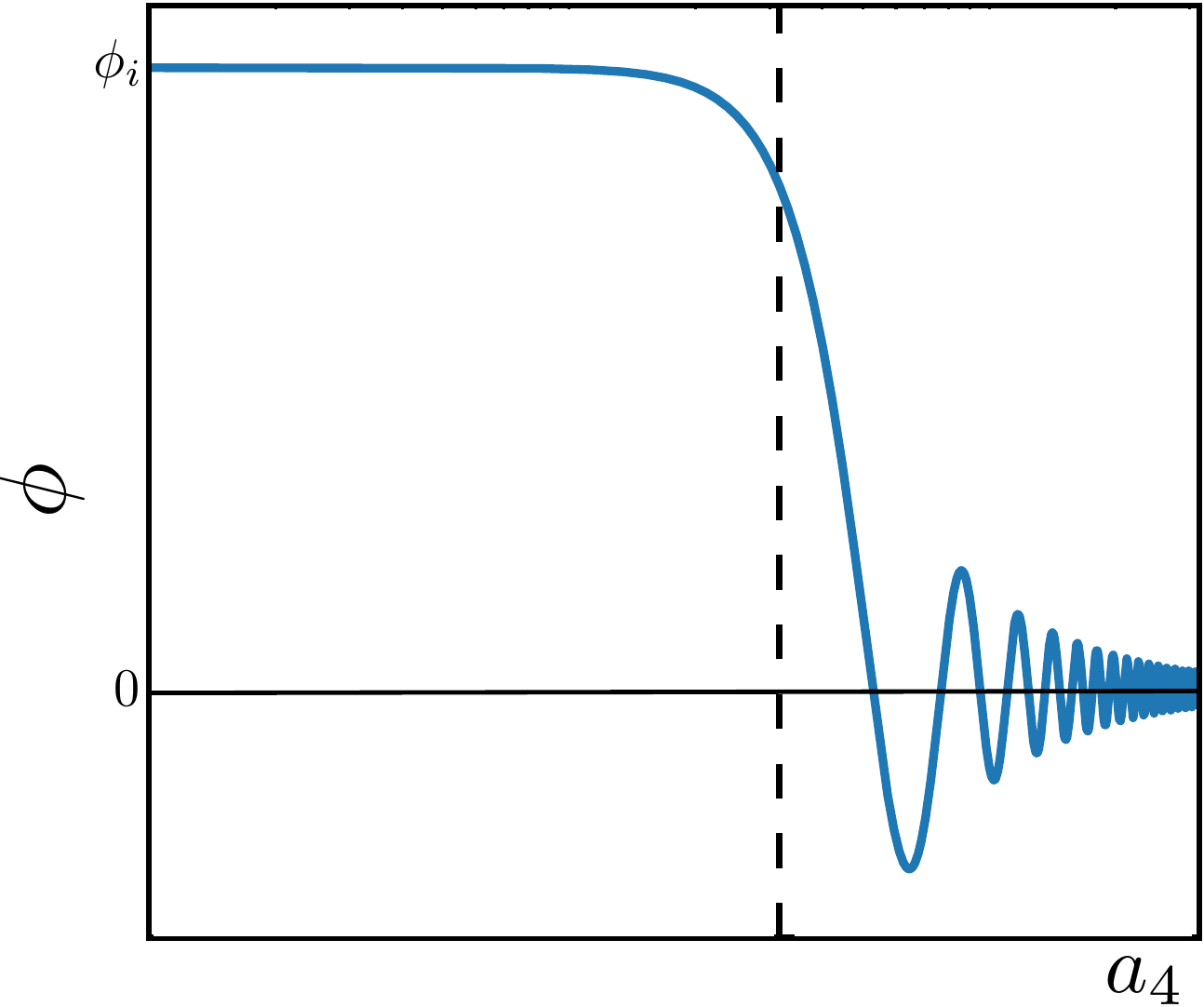}{0.92}
  \end{minipage}
  \begin{minipage}[t]{0.49\textwidth}
  \postscript{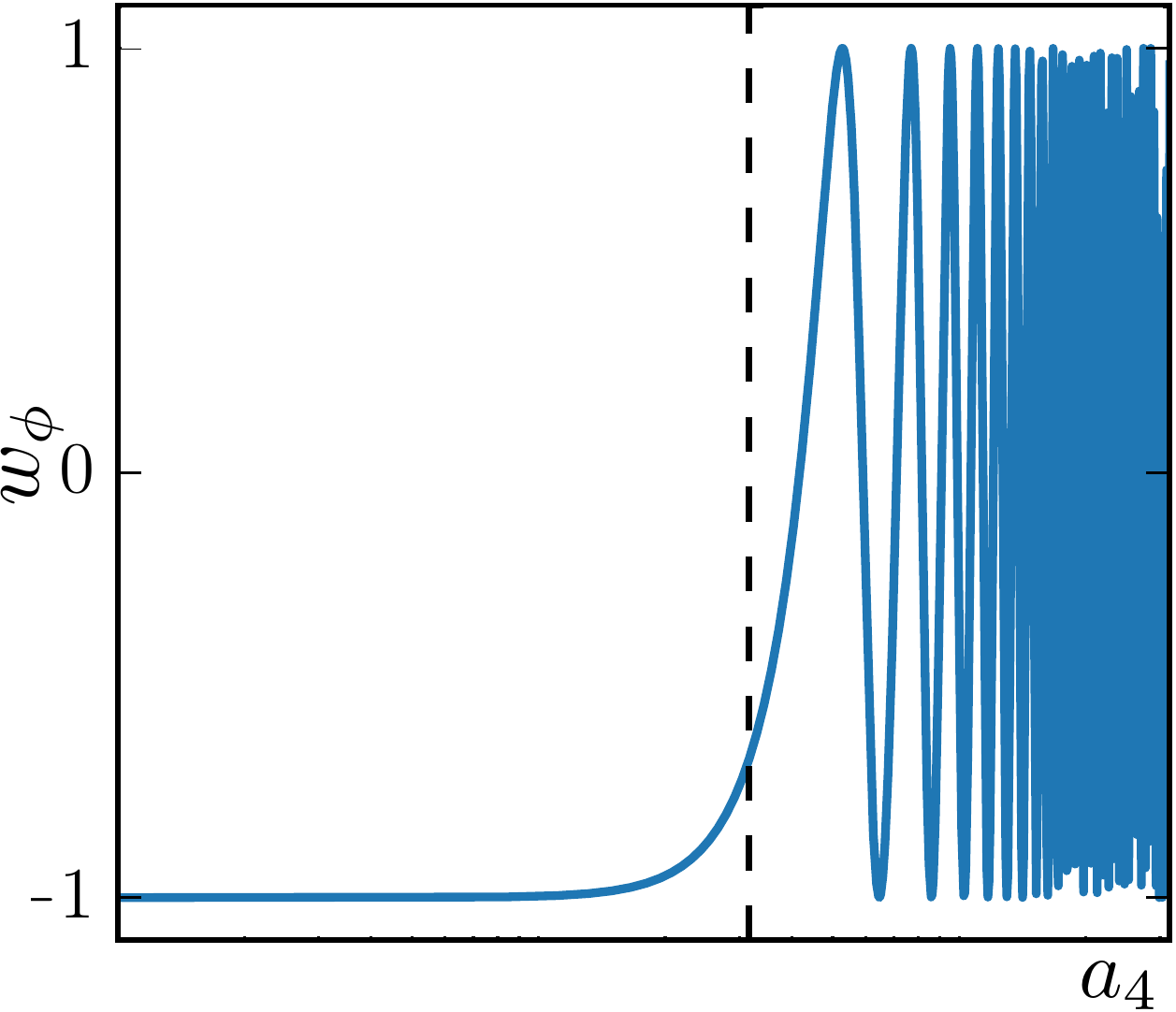}{0.9}
  \end{minipage}
  \caption{Schematic representation of the evolution of $\phi$ (left) and $w_\phi$ (right) with the
    scale factor $a_4$. The dimensionful quantities
    have arbitrary normalization. The vertical dashed lines indicate the
    condition defining $m \sim 3 H$. \label{fig:2}}
\end{figure*}

Qualitatively, radion cosmology reassembles that of ultralight axion-like
particles~\cite{Marsh:2015xka}. Namely, the radion equation of motion
is given by
\begin{equation}
  \ddot \phi + 3 H \dot \phi + V^{\rm tot}_\phi = 0 \,,
\label{RofM}
\end{equation}
where $H$ is the Hubble parameter. We assume that $\phi$ is around the
minimum of the potential at the origin such that the total
potential can be expanded around its minimum as $V^{\rm tot} \sim (m
\phi)^2/2 + \Lambda_4$, and so (\ref{RofM}) can be rewritten as
\begin{equation}
  \ddot \phi + 3 H \dot \phi + m^2 \phi = 0 \, .
\label{RofM2}
\end{equation}
At very early times, when $m<3H$, the radion field is overdamped and frozen at its
initial value by Hubble friction. During this epoch the equation of state 
is $w_\phi = -1$ and the radion behaves as a sub-dominant cosmological
constant. Once the Universe expands to the point where $m \sim 3H$, the
driving force overcomes the friction and the field begins to slowly
roll. Finally, when $m> 3H$, the field executes undamped oscillations. The equation of
state oscillates around $w_\phi = 0$ and the energy density scales as
CDM.  A visual representation of the evolution of $\phi$ and $w_\phi$
is shown in Fig.~\ref{fig:2}.

We now turn the estimate the required energy density of the radion
field $\rho_\phi$ to accommodate the observed relic density if the CDM evolution
 should duplicate that in $\Lambda$CDM after the epoch of matter-radiation
 equality. To this end we first reexamine the evolution of the radiation
energy density, which can be conveniently expressed as 
\begin{equation}
  \rho_R = \left(\sum_B g_B + \frac{7}{8} \sum_F g_F \right)
  \frac{\pi^2}{30} \ T^4 \equiv \frac{\pi^2}{30} \ N(T) \ T^4 \,,
\label{rhoR}
\end{equation}  
where $g_{B(F)}$ is the total number of boson (fermion) degrees of
freedom and the sum runs over all boson (fermion) states with $m_{B(F)} \ll 
T$, and where $N(T)$ is the number of effective degrees of freedom (the factor of $7/8$ is due to the difference between the Fermi and
Bose integrals).

The expansion rate as a function of the temperature in
the plasma is given by
\begin{equation}
  H (T) = \left[\frac{\pi^2}{90} N(T) \right]^{1/2} \frac{T^2}{M_p}
  \sim 0.33 \sqrt{N(T)} \, \frac{T^2}{M_p} \, . 
\label{HT}
\end{equation}
By inspection of (\ref{HT}) we can immediately see that for $m \sim 3H$, 
only photons and neutrinos contribute to the
sum in (\ref{rhoR}), yielding $N = 7.25$. This corresponds to a
temperature $T_{\rm osc} \sim
86~{\rm eV}$, for which the total energy density of radiation
(\ref{rhoR}) is $\rho_R (T_{\rm osc}) \sim 10^8~{\rm  
  eV}^4$. Now, let $\rho_\phi (T_{\rm osc})$ be the background energy
density of the radion field at $T_{\rm osc}$. As the
universe expands, the ratio of dark matter to radiation grows as
$1/T$, and in $\Lambda$CDM cosmology they are supposed to become
equal at the temperature $T_{\rm MR} \sim 1~{\rm eV}$ of
matter-radiation equality. This implies that \begin{equation}
  \frac{\rho_\phi (T_{\rm osc})}{\rho_R (T_{\rm osc})} \ \frac{T_{\rm
      osc}}{T_{\rm MR}} \sim 1 \,,
\end{equation}
which leads to $\rho_\phi (T_{\rm osc}) \sim 10^6~{\rm
  eV}^4$~\cite{Hui:2016ltb}. In other words, if the density of the radion
field were about $10^6~{\rm  eV^4}$, then today's radion abundance would
easily accommodate the observed dark matter density~\cite{Planck:2018vyg}, i.e., 
$\rho_{\phi, {\rm today}} \sim  \rho_{\rm DM} \sim 1.26~{\rm
  keV/cm^{3}}$. We note that $\rho_{\phi} (T_{\rm osc})$ should be
equal to the value of the potential (above $\Lambda_4$) at the
constant value $\phi_i$ that is the initial condition, i.e. $V^{\rm tot} (\phi_i) \sim
10^6~{\rm eV}^4$. We note that although the initial value of radion field is a free
parameter of the model it is subject to the constraint $\phi_i/M_p \ll
1$, so that $V^{\rm tot}_\phi
\sim m^2 \phi$ and the expansion in
(\ref{RofM2}) is valid.  

In closing, we note that when oscillations start, before the
matter-radiation equality, the radion is non-relativistic and
therefore $\Delta N_{\rm eff}$ (the number of ``equivalent'' light neutrino species in units of the density of a
single Weyl neutrino~\cite{Steigman:1977kc}) stays unaffected at the
earliest observationally verified landmarks ({\it viz.} big bang
nucleosynthesis and cosmic microwave background). As a consequence, our model
remains consistent with the bounds derived in~\cite{Cicoli:2012aq,Higaki:2012ar}.

\section{Conclusions}
\label{sec:4}

We have introduced a new dark matter contender within the context of the
dark dimension. The dramatis personae is the radion, a bulk scalar
field whose quintessence-like potential drives an inflationary phase 
described by a 5-dimensional de Sitter (or approximate) solution of
Einstein equations. We have shown that within this set up the radion could be ultralight and
thereby serve as a fuzzy dark matter candidate. We have put forward a
simple cosmological production mechanism bringing into play unstable
KK graviton towers which are fueled via inflaton decay.

We end with an observation. The coherent oscillation of the fuzzy
radion in galactic haloes leads to pressure perturbations oscillating
at twice its Compton frequency, $\omega =
2m$~\cite{Khmelnitsky:2013lxt}. These oscillations induce fluctuations of the
gravitational potential at frequency
\begin{eqnarray}
  f & \equiv & \omega/(2\pi) \nonumber \\
  & \simeq & 4.8 \times 10^{-9}~{\rm Hz} \ \left(\frac{m}{10^{-23}~{\rm eV}}\right)  
\end{eqnarray}
and can give rise to distinctive profiles in the travel time of radio beams
emitted from pulsars, which have been monitored for decades in Pulsar
Timing Array (PTA) experiments~\cite{Anholm:2008wy}.

Recently, several PTA experiments reported $4\sigma$ evidence for a
stochastic signal with quadrupolar angular correlations consistent with the expected
isotropic background of gravitational waves
radiated by inspiraling supermassive black hole
binaries~\cite{NANOGrav:2023gor,Reardon:2023gzh,Antoniadis:2023ott,Xu:2023wog}. (A
plethora of models have also been explored to
explain the Hellings-Downs-correlated signal~\cite{Vagnozzi:2023lwo,Megias:2023kiy,Ellis:2023tsl,Lazarides:2023ksx,Han:2023olf,Guo:2023hyp,Yang:2023aak}.)  In
addition, the NANOGrav Collaboration reported a monopole excess around
4~nHz~\cite{NANOGrav:2023hvm}. The fuzzy radion can accommodate the
PTA monopole anomaly if  $m \sim 8 \times 10^{-24}~{\rm eV}$.\\

\section*{Acknowledgments}

We have greatly benefited from discussions with Nima Arkani-Hamed. 
We thank Jules Cunat for discussion, and Andrea Mitridate and Sunny Vagnozzi for valuable email
correspondence about the PTA signal.  The research of LAA is supported
by the U.S. National Science Foundation (NSF grant PHY-2112527). The
work of D.L. is supported by the Origins Excellence Cluster and by the
German-Israel-Project (DIP) on Holography and the Swampland.


\begin{thebibliography}{99}

\bibitem{Feng:2010gw}
J.~L.~Feng,
{\color{rossoCP3} Dark Matter Candidates from Particle Physics and Methods of Detection},
Ann. Rev. Astron. Astrophys. \textbf{48}, 495-545 (2010)
doi:10.1146/annurev-astro-082708-101659
[arXiv:1003.0904 [astro-ph.CO]].
  
\bibitem{Hu:2000ke}
W.~Hu, R.~Barkana and A.~Gruzinov,
{\color{rossoCP3} Cold and fuzzy dark matter},
Phys. Rev. Lett. \textbf{85}, 1158-1161 (2000)
doi:10.1103/PhysRevLett.85.1158
[arXiv:astro-ph/0003365 [astro-ph]].






\bibitem{Irsic:2017yje}
V.~Ir\v{s}i\v{c}, M.~Viel, M.~G.~Haehnelt, J.~S.~Bolton and G.~D.~Becker,
{\color{rossoCP3} First constraints on fuzzy dark matter from Lyman-$\alpha$ forest data and hydrodynamical simulations},
Phys. Rev. Lett. \textbf{119}, no.3, 031302 (2017)
doi:10.1103/PhysRevLett.119.031302
[arXiv:1703.04683 [astro-ph.CO]].

\bibitem{Armengaud:2017nkf}
E.~Armengaud, N.~Palanque-Delabrouille, C.~Y\`eche, D.~J.~E.~Marsh and J.~Baur,
{\color{rossoCP3} Constraining the mass of light bosonic dark matter using SDSS Lyman-$\alpha$ forest},
Mon. Not. Roy. Astron. Soc. \textbf{471}, no.4, 4606-4614 (2017)
doi:10.1093/mnras/stx1870
[arXiv:1703.09126 [astro-ph.CO]].

\bibitem{Rogers:2020ltq}
K.~K.~Rogers and H.~V.~Peiris,
{\color{rossoCP3} Strong Bound on Canonical Ultralight Axion Dark Matter from the Lyman-Alpha Forest},
Phys. Rev. Lett. \textbf{126}, no.7, 071302 (2021)
doi:10.1103/PhysRevLett.126.071302
[arXiv:2007.12705 [astro-ph.CO]].


\bibitem{Hui:2016ltb}
L.~Hui, J.~P.~Ostriker, S.~Tremaine and E.~Witten,
{\color{rossoCP3} Ultralight scalars as cosmological dark matter},
Phys. Rev. D \textbf{95}, no.4, 043541 (2017)
doi:10.1103/PhysRevD.95.043541
[arXiv:1610.08297 [astro-ph.CO]].

\bibitem{Schutz:2020jox}
K.~Schutz,
{\color{rossoCP3}  Subhalo mass function and ultralight bosonic dark matter},
Phys. Rev. D \textbf{101} (2020) no.12, 123026
doi:10.1103/PhysRevD.101.123026
[arXiv:2001.05503 [astro-ph.CO]].

\bibitem{Dalal:2022rmp}
N.~Dalal and A.~Kravtsov,
{\color{rossoCP3}  Excluding fuzzy dark matter with sizes and stellar kinematics of ultrafaint dwarf galaxies},
Phys. Rev. D \textbf{106} (2022) no.6, 063517
doi:10.1103/PhysRevD.106.063517
[arXiv:2203.05750 [astro-ph.CO]].

\bibitem{Davoudiasl:2019nlo}
H.~Davoudiasl and P.~B.~Denton,
{\color{rossoCP3}  Ultralight boson dark matter and Event Horizon Telescope observations of M87$^*$},
Phys. Rev. Lett. \textbf{123} (2019) no.2, 021102
doi:10.1103/PhysRevLett.123.021102
[arXiv:1904.09242 [astro-ph.CO]].



\bibitem{Montero:2022prj}
M.~Montero, C.~Vafa and I.~Valenzuela,
{\color{rossoCP3} The dark dimension and the Swampland},
JHEP \textbf{02}, 022 (2023)
doi:10.1007/JHEP02(2023)022
[arXiv:2205.12293 [hep-th]].


\bibitem{Vafa:2005ui}
C.~Vafa,
{\color{rossoCP3} The string landscape and the swampland},
[arXiv:hep-th/0509212 [hep-th]].


\bibitem{vanBeest:2021lhn}
M.~van Beest, J.~Calder\'on-Infante, D.~Mirfendereski and I.~Valenzuela,
{\color{rossoCP3} Lectures on the Swampland Program in string compactifications},
Phys. Rept. \textbf{989}, 1-50 (2022)
doi:10.1016/j.physrep.2022.09.002
[arXiv:2102.01111 [hep-th]].

\bibitem{Palti:2019pca}
E.~Palti,
{\color{rossoCP3} The swampland: introduction and review},
Fortsch. Phys. \textbf{67}, no.6, 1900037 (2019)
doi:10.1002/prop.201900037
[arXiv:1903.06239 [hep-th]].

\bibitem{Agmon:2022thq}
N.~B.~Agmon, A.~Bedroya, M.~J.~Kang and C.~Vafa,
{\color{rossoCP3} Lectures on the string landscape and the swampland},
[arXiv:2212.06187 [hep-th]].


\bibitem{Ooguri:2006in}
H.~Ooguri and C.~Vafa,
{\color{rossoCP3} On the Geometry of the String Landscape and the Swampland},
Nucl. Phys. B \textbf{766}, 21-33 (2007)
doi:10.1016/j.nuclphysb.2006.10.033
[arXiv:hep-th/0605264 [hep-th]].

\bibitem{Lust:2019zwm}
D.~L\"ust, E.~Palti and C.~Vafa,
 {\color{rossoCP3} AdS and the Swampland},
Phys. Lett. B \textbf{797}, 134867 (2019)
doi:10.1016/j.physletb.2019.134867
[arXiv:1906.05225 [hep-th]].

\bibitem{Higuchi:1986py}
A.~Higuchi,
 {\color{rossoCP3} Forbidden mass range for spin-2 field theory in de Sitter space-time},
Nucl. Phys. B \textbf{282} (1987), 397-436
doi:10.1016/0550-3213(87)90691-2


\bibitem{Itoyama:1986ei}
H.~Itoyama and T.~R.~Taylor,
 {\color{rossoCP3} Supersymmetry restoration in the compactified $O(16) \times O(16)$-prime heterotic string theory},
Phys. Lett. B \textbf{186} (1987), 129-133
doi:10.1016/0370-2693(87)90267-X

\bibitem{Itoyama:1987rc}
H.~Itoyama and T.~R.~Taylor,
 {\color{rossoCP3} Small cosmological constant in string models},
FERMILAB-CONF-87-129-T.

\bibitem{Antoniadis:1991kh}
I.~Antoniadis and C.~Kounnas,
 {\color{rossoCP3} Superstring phase transition at high temperature},
Phys. Lett. B \textbf{261} (1991), 369-378
doi:10.1016/0370-2693(91)90442-S





\bibitem{Bonnefoy:2018tcp}
Q.~Bonnefoy, E.~Dudas and S.~L\"ust,
 {\color{rossoCP3}  On the weak gravity conjecture in string theory with broken supersymmetry},
Nucl. Phys. B \textbf{947} (2019), 114738
doi:10.1016/j.nuclphysb.2019.114738
[arXiv:1811.11199 [hep-th]].




\bibitem{Anchordoqui:2023laz}
L.~A.~Anchordoqui, I.~Antoniadis, D.~L\"ust and S.~L\"ust,
 {\color{rossoCP3} On the cosmological constant, the KK mass scale, and the cut-off dependence in the dark dimension scenario},
[arXiv:2309.09330 [hep-th]].

\bibitem{Lee:2020zjt}
J.~G.~Lee, E.~G.~Adelberger, T.~S.~Cook, S.~M.~Fleischer and B.~R.~Heckel,
 {\color{rossoCP3}  New test of the gravitational $1/r^2$ law at separations down to 52 $\mu$m},
Phys. Rev. Lett. \textbf{124}, no.10, 101101 (2020)
doi:10.1103/PhysRevLett.124.101101
[arXiv:2002.11761 [hep-ex]].

\bibitem{Hannestad:2003yd}
S.~Hannestad and G.~G.~Raffelt,
 {\color{rossoCP3} Supernova and neutron star limits on large extra dimensions reexamined},
Phys. Rev. D \textbf{67}, 125008 (2003)
[erratum: Phys. Rev. D \textbf{69}, 029901 (2004)]
doi:10.1103/PhysRevD.69.029901
[arXiv:hep-ph/0304029 [hep-ph]].





\bibitem{Dvali:2007hz}
G.~Dvali,
{\color{rossoCP3} Black holes and large $N$ species solution to the hierarchy problem},
Fortsch. Phys. \textbf{58}, 528-536 (2010)
doi:10.1002/prop.201000009
[arXiv:0706.2050 [hep-th]].

\bibitem{Dvali:2007wp}
G.~Dvali and M.~Redi,
{\color{rossoCP3} Black hole bound on the number of species and quantum gravity at LHC},
Phys. Rev. D \textbf{77}, 045027 (2008)
doi:10.1103/PhysRevD.77.045027
[arXiv:0710.4344 [hep-th]].

\bibitem{ParticleDataGroup:2022pth}
R.~L.~Workman \textit{et al.} [Particle Data Group],
{\color{rossoCP3} Review of Particle Physics},
PTEP \textbf{2022}, 083C01 (2022)
doi:10.1093/ptep/ptac097


\bibitem{Anchordoqui:2022ejw}
L.~A.~Anchordoqui,
{\color{rossoCP3} Dark dimension, the swampland, and the origin of cosmic rays beyond the Greisen-Zatsepin-Kuzmin barrier},
Phys. Rev. D \textbf{106}, no.11, 116022 (2022)
doi:10.1103/PhysRevD.106.116022
[arXiv:2205.13931 [hep-ph]].


\bibitem{Anchordoqui:2022txe}
L.~A.~Anchordoqui, I.~Antoniadis and D.~L\"ust,
 {\color{rossoCP3} Dark dimension, the swampland, and the dark matter fraction composed of primordial black holes},
Phys. Rev. D \textbf{106}, no.8, 086001 (2022)
doi:10.1103/PhysRevD.106.086001
[arXiv:2206.07071 [hep-th]].

\bibitem{Blumenhagen:2022zzw}
R.~Blumenhagen, M.~Brinkmann and A.~Makridou,
 {\color{rossoCP3}  The dark dimension in a warped throat},
Phys. Lett. B \textbf{838}, 137699 (2023)
doi:10.1016/j.physletb.2023.137699
[arXiv:2208.01057 [hep-th]].


\bibitem{Anchordoqui:2022tgp}
L.~A.~Anchordoqui, I.~Antoniadis and D.~L\"ust,
 {\color{rossoCP3} The dark universe: Primordial black hole \ensuremath{\leftrightharpoons} dark graviton gas connection},
Phys. Lett. B \textbf{840}, 137844 (2023)
doi:10.1016/j.physletb.2023.137844
[arXiv:2210.02475 [hep-th]].

\bibitem{Gonzalo:2022jac}
E.~Gonzalo, M.~Montero, G.~Obied and C.~Vafa,
 {\color{rossoCP3} Dark Dimension Gravitons as Dark Matter},
[arXiv:2209.09249 [hep-ph]].


\bibitem{Anchordoqui:2022svl}
L.~A.~Anchordoqui, I.~Antoniadis and D.~L\"ust,
 {\color{rossoCP3} Aspects of the dark dimension in cosmology},
Phys. Rev. D \textbf{107}, no.8, 083530 (2023)
doi:10.1103/PhysRevD.107.083530
[arXiv:2212.08527 [hep-ph]].





\bibitem{Anchordoqui:2023oqm}
L.~A.~Anchordoqui, I.~Antoniadis, N.~Cribiori, D.~L\"ust and M.~Scalisi,
 {\color{rossoCP3} The Scale of Supersymmetry Breaking and the Dark Dimension},
JHEP \textbf{05}, 060 (2023)
doi:10.1007/JHEP05(2023)060
[arXiv:2301.07719 [hep-th]].




\bibitem{vandeHeisteeg:2023uxj}
D.~van de Heisteeg, C.~Vafa, M.~Wiesner and D.~H.~Wu,
 {\color{rossoCP3} Bounds on Field Range for Slowly Varying Positive Potentials},
[arXiv:2305.07701 [hep-th]].

\bibitem{Noble:2023mfw}
N.~T.~Noble, J.~F.~Soriano and L.~A.~Anchordoqui,
{\color{rossoCP3}  Probing the Dark Dimension with Auger data},
Phys. Dark Univ. (in press)
[arXiv:2306.03666 [hep-ph]].


\bibitem{Anchordoqui:2023wkm}
L.~A.~Anchordoqui, I.~Antoniadis and J.~Cunat,
 {\color{rossoCP3} The Dark Dimension and the Standard Model Landscape},
[arXiv:2306.16491 [hep-ph]].

\bibitem{Anchordoqui:2023qxv}
L.~A.~Anchordoqui, I.~Antoniadis, K.~Benakli, J.~Cunat and D.~L\"ust,
 {\color{rossoCP3}  Searching for neutrino-modulino oscillations at the Forward Physics Facility},
[arXiv:2308.11476 [hep-ph]].

\bibitem{Anchordoqui:2023etp}
L.~A.~Anchordoqui and I.~Antoniadis,
 {\color{rossoCP3}  Large extra dimensions from higher-dimensional inflation},
[arXiv:2310.20282 [hep-ph]].



\bibitem{Barreiro:1999zs}
T.~Barreiro, E.~J.~Copeland and N.~J.~Nunes,
 {\color{rossoCP3}  Quintessence arising from exponential potentials},
Phys. Rev. D \textbf{61}, 127301 (2000)
doi:10.1103/PhysRevD.61.127301
[arXiv:astro-ph/9910214 [astro-ph]].




\bibitem{Obied:2018sgi}
G.~Obied, H.~Ooguri, L.~Spodyneiko and C.~Vafa,
 {\color{rossoCP3}  De Sitter Space and the Swampland},
[arXiv:1806.08362 [hep-th]].

\bibitem{Arkani-Hamed:2007ryu}
N.~Arkani-Hamed, S.~Dubovsky, A.~Nicolis and G.~Villadoro,
 {\color{rossoCP3}  Quantum horizons of the Standard Model landscape},
JHEP \textbf{06}, 078 (2007)
doi:10.1088/1126-6708/2007/06/078
[arXiv:hep-th/0703067 [hep-th]].

  

\bibitem{Albrecht:2001cp}
A.~Albrecht, C.~P.~Burgess, F.~Ravndal and C.~Skordis,
 {\color{rossoCP3} Exponentially large extra dimensions},
Phys. Rev. D \textbf{65}, 123506 (2002)
doi:10.1103/PhysRevD.65.123506
[arXiv:hep-th/0105261 [hep-th]].






\bibitem{Dienes:2011ja}
K.~R.~Dienes and B.~Thomas,
 {\color{rossoCP3}  Dynamical Dark Matter: I. Theoretical Overview},
Phys. Rev. D \textbf{85}, 083523 (2012)
doi:10.1103/PhysRevD.85.083523
[arXiv:1106.4546 [hep-ph]].


\bibitem{Mohapatra:2003ah}
R.~N.~Mohapatra, S.~Nussinov and A.~Perez-Lorenzana,
 {\color{rossoCP3} Large extra dimensions and decaying $K K$ recurrences},
Phys. Rev. D \textbf{68}, 116001 (2003)
doi:10.1103/PhysRevD.68.116001
[arXiv:hep-ph/0308051 [hep-ph]].


\bibitem{Dienes:2008qi}
K.~R.~Dienes and B.~Thomas,
 {\color{rossoCP3}  Cascades and Collapses, Great Walls and Forbidden Cities: Infinite Towers of Metastable Vacua in Supersymmetric Field Theories},
Phys. Rev. D \textbf{79}, 045001 (2009)
doi:10.1103/PhysRevD.79.045001
[arXiv:0811.3335 [hep-th]].


\bibitem{Marsh:2015xka}
D.~J.~E.~Marsh,
 {\color{rossoCP3}  Axion Cosmology},
Phys. Rept. \textbf{643}, 1-79 (2016)
doi:10.1016/j.physrep.2016.06.005
[arXiv:1510.07633 [astro-ph.CO]].



\bibitem{Planck:2018vyg}
N.~Aghanim \textit{et al.} [Planck],
 {\color{rossoCP3}  Planck 2018 results. VI. Cosmological parameters},
Astron. Astrophys. \textbf{641}, A6 (2020)
[erratum: Astron. Astrophys. \textbf{652}, C4 (2021)]
doi:10.1051/0004-6361/201833910
[arXiv:1807.06209 [astro-ph.CO]].

\bibitem{Steigman:1977kc}
G.~Steigman, D.~N.~Schramm and J.~E.~Gunn,
  {\color{rossoCP3}  Cosmological Limits to the Number of Massive Leptons},
Phys. Lett. B \textbf{66}, 202-204 (1977)
doi:10.1016/0370-2693(77)90176-9


\bibitem{Cicoli:2012aq}
M.~Cicoli, J.~P.~Conlon and F.~Quevedo,
 {\color{rossoCP3}  Dark radiation in LARGE volume models},
Phys. Rev. D \textbf{87}, no.4, 043520 (2013)
doi:10.1103/PhysRevD.87.043520
[arXiv:1208.3562 [hep-ph]].


\bibitem{Higaki:2012ar}
T.~Higaki and F.~Takahashi,
 {\color{rossoCP3}  Dark Radiation and Dark Matter in Large Volume Compactifications},
JHEP \textbf{11}, 125 (2012)
doi:10.1007/JHEP11(2012)125
[arXiv:1208.3563 [hep-ph]].



\bibitem{Khmelnitsky:2013lxt}
A.~Khmelnitsky and V.~Rubakov,
{\color{rossoCP3} Pulsar timing signal from ultralight scalar dark matter},
JCAP \textbf{02}, 019 (2014)
doi:10.1088/1475-7516/2014/02/019
[arXiv:1309.5888 [astro-ph.CO]].







\bibitem{Anholm:2008wy}
M.~Anholm, S.~Ballmer, J.~D.~E.~Creighton, L.~R.~Price and X.~Siemens,
{\color{rossoCP3} Optimal strategies for gravitational wave stochastic background searches in pulsar timing data},
Phys. Rev. D \textbf{79}, 084030 (2009)
doi:10.1103/PhysRevD.79.084030
[arXiv:0809.0701 [gr-qc]].


\bibitem{NANOGrav:2023gor}
G.~Agazie \textit{et al.} [NANOGrav],
{\color{rossoCP3} The NANOGrav 15-year Data Set: Evidence for a Gravitational-Wave Background},
doi:10.3847/2041-8213/acdac6
[arXiv:2306.16213 [astro-ph.HE]].



\bibitem{Reardon:2023gzh}
D.~J.~Reardon, A.~Zic, R.~M.~Shannon, G.~B.~Hobbs, M.~Bailes, V.~Di Marco, A.~Kapur, A.~F.~Rogers, E.~Thrane and J.~Askew, \textit{et al.}
{\color{rossoCP3} Search for an isotropic gravitational-wave background with the Parkes Pulsar Timing Array},
doi:10.3847/2041-8213/acdd02
[arXiv:2306.16215 [astro-ph.HE]].


\bibitem{Antoniadis:2023ott}
J.~Antoniadis, P.~Arumugam, S.~Arumugam, S.~Babak, M.~Bagchi, A.~S.~B.~Nielsen, C.~G.~Bassa, A.~Bathula, A.~Berthereau and M.~Bonetti, \textit{et al.}
{\color{rossoCP3} The second data release from the European Pulsar Timing Array III. Search for gravitational wave signals},
[arXiv:2306.16214 [astro-ph.HE]].




\bibitem{Xu:2023wog}
H.~Xu, S.~Chen, Y.~Guo, J.~Jiang, B.~Wang, J.~Xu, Z.~Xue, R.~N.~Caballero, J.~Yuan and Y.~Xu, \textit{et al.}
{\color{rossoCP3} Searching for the nano-Hertz stochastic gravitational wave background with the Chinese Pulsar Timing Array Data Release I},
doi:10.1088/1674-4527/acdfa5
[arXiv:2306.16216 [astro-ph.HE]].

\bibitem{Vagnozzi:2023lwo}
S.~Vagnozzi,
{\color{rossoCP3} Inflationary interpretation of the stochastic gravitational wave background signal detected by pulsar timing array experiments},
[arXiv:2306.16912 [astro-ph.CO]].

\bibitem{Megias:2023kiy}
E.~Megias, G.~Nardini and M.~Quiros,
{\color{rossoCP3}  Pulsar Timing Array Stochastic Background from light Kaluza-Klein resonances},
[arXiv:2306.17071 [hep-ph]].


\bibitem{Ellis:2023tsl}
J.~Ellis, M.~Lewicki, C.~Lin and V.~Vaskonen,
 {\color{rossoCP3}  Cosmic Superstrings Revisited in Light of NANOGrav 15-Year Data},
[arXiv:2306.17147 [astro-ph.CO]].




\bibitem{Lazarides:2023ksx}
G.~Lazarides, R.~Maji and Q.~Shafi,
{\color{rossoCP3}   Superheavy quasi-stable strings and walls bounded by strings in the light of NANOGrav 15 year data},
[arXiv:2306.17788 [hep-ph]].

\bibitem{Han:2023olf}
C.~Han, K.~P.~Xie, J.~M.~Yang and M.~Zhang,
{\color{rossoCP3}  Self-interacting dark matter implied by nano-Hertz gravitational waves},
[arXiv:2306.16966 [hep-ph]].

\bibitem{Guo:2023hyp}
S.~Y.~Guo, M.~Khlopov, X.~Liu, L.~Wu, Y.~Wu and B.~Zhu,
{\color{rossoCP3}  Footprints of Axion-Like Particle in Pulsar Timing Array Data and JWST Observations},
[arXiv:2306.17022 [hep-ph]].

\bibitem{Yang:2023aak}
J.~Yang, N.~Xie and F.~P.~Huang,
{\color{rossoCP3}  Nano-Hertz stochastic gravitational wave background as hints of ultralight axion particles},
[arXiv:2306.17113 [hep-ph]].





\bibitem{NANOGrav:2023hvm}
A.~Afzal \textit{et al.} [NANOGrav],
{\color{rossoCP3} The NANOGrav 15-year Data Set: Search for Signals from New Physics},
Astrophys. J. Lett. \textbf{951}, no.1, L11 (2023)
doi:10.3847/2041-8213/acdc91
[arXiv:2306.16219 [astro-ph.HE]].



\end{thebibliography}
\end{document}